\documentclass[cameraready]{Interspeech}
\usepackage[table,xcdraw]{xcolor}
\usepackage{multirow}
\usepackage{caption}
\usepackage{subcaption}
\usepackage{cite}
\title{An Evaluation Framework for Text-to-Speech Voice Reconstruction}

\author[orcid=0009-0000-4721-9147, equalcontribution, correspondingauthor]{Ariadna}{Sanchez}
\author[orcid=0009-0007-8990-3941, equalcontribution, correspondingauthor]{Christoph}{Minixhofer}
\author[orcid=0000-0003-1450-8270]{Korin}{Richmond} 
\author[orcid=0000-0001-5495-967X]{Ond\v{r}ej}{Klejch}
\author[]{\\Peter}{Bell}
\author[orcid=0000-0002-2694-2843]{Simon}{King}

\address{
    The Centre for Speech Technology Research, University of Edinburgh, UK
}

\email{ariadna.sanchez@ed.ac.uk, christoph.minixhofer@ed.ac.uk}


\usepackage{comment}

\begin{document}

\maketitle

\begin{abstract}
Voice reconstruction using Text-to-Speech (TTS) offers a communication method for people with speech disorders, which aims to retain their speaker identity while improving intelligibility. Previous work generally relies on Mean Opinion Score (MOS) to evaluate naturalness and speaker similarity, but this has limited sensitivity and reliability. We propose an evaluation framework with subjective and objective components. Subjectively, we evaluate perceived intelligibility and speaker identity using Best Worst Scaling (BWS) with situational framing. Objectively, we demonstrate that standard measures fail to predict reconstruction success for highly unintelligible speakers, so we introduce a novel dual-reference distributional measure to assess the trade-off between intelligibility and speaker identity. By evaluating the output of 17 zero-shot TTS systems for 193 speakers, we show that our framework provides a reliable and task-aligned approach for assessing voice reconstruction.
\end{abstract}

\section{Introduction}

Speech disorders, caused by neurological conditions, can impact someone's ability to communicate. Eventually, some of these people will become users of a voice output communication aid (VOCA), which often uses Text-to-Speech (TTS). Voice reconstruction is the task of creating a personalised TTS voice for speakers whose speech is already disordered, maintaining their speaker identity while improving the intelligibility of their speech.
Even though current TTS technologies have achieved high fidelity when replicating people with typical speech, voice reconstruction still remains challenging due to the difficulty of disentangling desired speaker characteristics from articulation patterns caused by the speaker's condition.

Previous work on voice reconstruction has mostly been evaluated using common TTS evaluation processes, i.e., evaluating naturalness and speaker similarity in listening tests using Mean Opinion Score (MOS) (e.g., \cite{veaux2015comparison, tian2024creating, jeon2025facilitating}). However, MOS is now known to have limitations including sensitivity, reliability, and saturation across studies \cite{le2024limits, varadhan2025the}, leading to calls for a shift in TTS evaluation \cite{bailly2025hot} towards use-case specific evaluation rather than generalised "naturalness" scores \cite{cooper2025good}. For voice reconstruction in particular, evaluation is difficult because no ground-truth data is available: what a speakers' reconstructed voice should sound like is subjective.

Common TTS evaluation protocols feature a blend of subjective and objective methods \cite{cooper2024review}. Common objective evaluations of voice reconstruction measure: Word Error Rate (WER), Character Error Rate (CER), or Phone Error Rate (PER), as proxies for intelligibility; cosine similarity of speaker embeddings, as a proxy for speaker similarity. Inspired by distributional evaluation in computer vision \cite{heusel2017fid}, recent work has applied this paradigm to objective evaluation of speech and audio \cite{kilgour2019fad} and shown correlation with listener preferences \cite{minixhofer2024ttsds, minixhofer2025ttsds2, cooper2025layer}. This approach is difficult to apply to voice reconstruction due to the lack of pre-condition reference data \cite{cooper2025layer,cooper2025progress}. Despite widespread use, the ability of objective measures to predict listener preferences for voice reconstruction is unexplored.

In previous work, TTS systems trained solely on typical speech have been used for the voice reconstruction task by providing a disordered speech prompt at inference time\cite{azizah2024zero, szekely2025voice}. To validate our evaluation framework with sufficient data, we use this approach, testing 17 zero-shot voice-cloning TTS systems.

For subjective evaluation, we adopt Best Worst Scaling (BWS) \cite{wells24_interspeech}, which we present to listeners in two situationally-framed tasks \cite{edlund2024assessing}: one assessing intelligibility without regard to speaker characteristics (\textsc{intelligibility}), and the other assessing overall reconstruction with respect to both intelligibility and speaker characteristics (\textsc{reconstruction}).
For objective evaluation, we investigate the correlation of measures estimated using Automatic Speech Recognition (ASR) models \cite{radford2023whisper,li2020allosaurus} (WER and PER), a speaker verification model \cite{wang2023wespeaker} (speaker cosine similarity), and a MOS prediction model\cite{saeki22c_interspeech} (UTMOS) with system rankings obtained from the listening tests. 

Voice reconstruction can be seen as the task of producing speech with higher intelligibility but perceptually similar speaker identity for a given reference. With distributional evaluation, the former can be approximated by computing the distance to high-intelligibility speech with differing speaker identity, while the latter can be approximated using the distance to the original disordered speech. A good voice reconstruction system should manage the trade-off between minimising both distances simultaneously. We specifically use TTSDS2 \cite{minixhofer2025ttsds2}, a distributional measure which was shown to correlate with listener preferences. We average the similarities of the synthesised speech to a generic high-intelligibility corpus (for intelligibility), and the original recordings (for speaker identity). All correlation results, audio examples and listening test instructions can be found at our project page\footnote{\url{https://minixc.github.io/sap/}}.

\section{Related work}

TTS voice reconstruction has mostly been evaluated for naturalness, intelligibility, and similarity to the target speaker. 
Most previous work evaluates \textbf{naturalness} with MOS \cite{tian2024creating, azizah2024zero, jeon2025facilitating}. Recently, \cite{szekely2025voice} proposed evaluating suitability of TTS output for human conversation, using a MUSHRA listening test. Automatic MOS prediction models such as UTMOS \cite{saeki22c_interspeech} have also been used as a proxy for listeners in the voice reconstruction task \cite{pu2025reinforcement}. Automatic MOS prediction remains challenging for out-of-domain samples such as disordered speech \cite{cooper2025progress, sanchez25_interspeech}.

How \textbf{intelligibility} is measured has shifted from reporting human transcription accuracy \cite{veaux2012using, veaux2015comparison} to relying on ASR to report WER \cite{wagner25_interspeech, sanchez25_interspeech, rosero2025finding}, CER \cite{azizah2024zero}, or PER \cite{tian2024creating, chen25m_interspeech, jeon2025facilitating} as a proxy of intelligibility. Expert speech and language therapists have also been used in previous work\cite{rosero2025finding, szekely2025voice}.

Evaluations for \textbf{similarity} to the target speaker have relied on a combination of subjective and objective evaluations. \cite{veaux2012using, veaux2015comparison, tian2024creating, jeon2025facilitating} measured speaker similarity with subjective ratings in a 5-point scale, commonly known as Similarity Mean Opinion Score (SMOS). Other work has measured speaker similarity by calculating cosine similarity of speaker embeddings from the target speaker to speaker embeddings of the synthetic output \cite{azizah2024zero, jeon2025facilitating, szekely2025voice, sanchez25_interspeech, rosero2025finding}. Authors in \cite{chen24t_interspeech} measure speaker similarity with a speaker verification system instead.

Most of the previous work reviewed avoids situationally framing listeners, i.e., explaining the intended use of the synthetic speech, with the exception of \cite{szekely2025voice}. Yet \cite{edlund2024assessing} demonstrate that framing listeners with different tasks significantly changes how they evaluate a voice, highlighting the importance of providing detailed information to listeners.

In objective TTS evaluation of healthy speech, distances between distributions of samples along various latent spaces \cite{kilgour2019fad,minixhofer2024ttsds,minixhofer2025ttsds2} or model layers \cite{cooper2025layer} has shown promising results. Specifically, TTSDS2 \cite{minixhofer2025ttsds2}, which provides an overall score between 0 and 100 based on the distributional distance between a reference dataset and synthetic data, has been shown to correlate with listener preferences, even for data outside the domain of most TTS system's training data, such as children's speech.

\section{Dataset}
We selected a total of 193 speakers from the Speech Accessibility Project (SAP) dataset (December 2024 release) \cite{hasegawa2024community} with four different types of condition: Parkinson's, Cerebral Palsy, ALS, or Down Syndrome. We only kept speakers whose first language was English to avoid adding foreign accents as an additional confound, as TTS systems have shown to underperform in those \cite{zhong2025accentbox}. Table \ref{tab:total_speakers} summarises the speakers for which we generated stimuli. We divide speakers by condition and intelligibility - considered high for those with a Word Error Rate (WER) below 30\%, and low for WER equal or above 30\%. Speaker WER was calculated with 10 samples per speaker, using Whisper \cite{radford2023whisper} (\texttt{turbo} checkpoint). The speaker set is skewed towards Parkinson's (72\%) and high intelligibility (77.2\%).

\begin{table}[]
\centering
\caption{Breakdown of the speakers selected from the Speech Accessibility Project dataset by condition and intelligibility.} 
\begin{tabular}{
>{\columncolor[HTML]{FFFFFF}}l 
>{\columncolor[HTML]{FFFFFF}}c 
>{\columncolor[HTML]{FFFFFF}}c }
\hline
\multicolumn{1}{c}{\cellcolor[HTML]{FFFFFF}\textbf{Condition}} & \textbf{Intelligibility} & \textbf{Total speakers} \\ \hline
Amyotrophic Lateral                                            & High                     & 12                      \\
Sclerosis                                                      & Low                      & 5                       \\ \hline
Cerebral Palsy                                                 & High                     & 10                      \\
                                                               & Low                      & 20                      \\ \hline
Down Syndrome                                                  & High                     & 5                       \\
                                                               & Low                      & 2                       \\ \hline
Parkinson's                                            & High                     & 122                     \\
                                                               & Low                      & 17                      \\ \hline
\end{tabular}
\label{tab:total_speakers}
\end{table}

For each selected speaker, two utterances are chosen at random, one of which is used as reference for the zero-shot voice cloning systems, while only the transcript of the other is used. Using these, we generate 193 synthetic utterances (one per speaker) for each of the 17 selected TTS systems \cite{zhou2025indextts2,hu2026qwen3,eskimez2024e2,liao2024fish,chen2025f5,wang2024maskgct,pengvibevoice,peng2024voicecraft,gptsovits,lee2022hierspeech,li2023styletts,betker2023better,zhang2025vevo,metavoice,casanova2024xtts,whisperspeech,qin2023openvoice}, which are listed in Table \ref{tab:main_results}.
The systems were selected based on availability of model weights and to cover a wide range of architectures (e.g. auto- and non-auto-regressive) and training data (read speech, spontaneous speech or a mixture of both). While future work could systematically investigate potential links between these attributes and voice reconstruction performance, we focus on establishing an evaluation protocol which can reliably identify differences between systems which are relevant to the voice reconstruction task, as outlined in the next section.

\begin{figure*}[t]
    \centering
    \begin{subfigure}[b]{0.49\textwidth}
        \centering
        \includegraphics[width=0.925\textwidth]{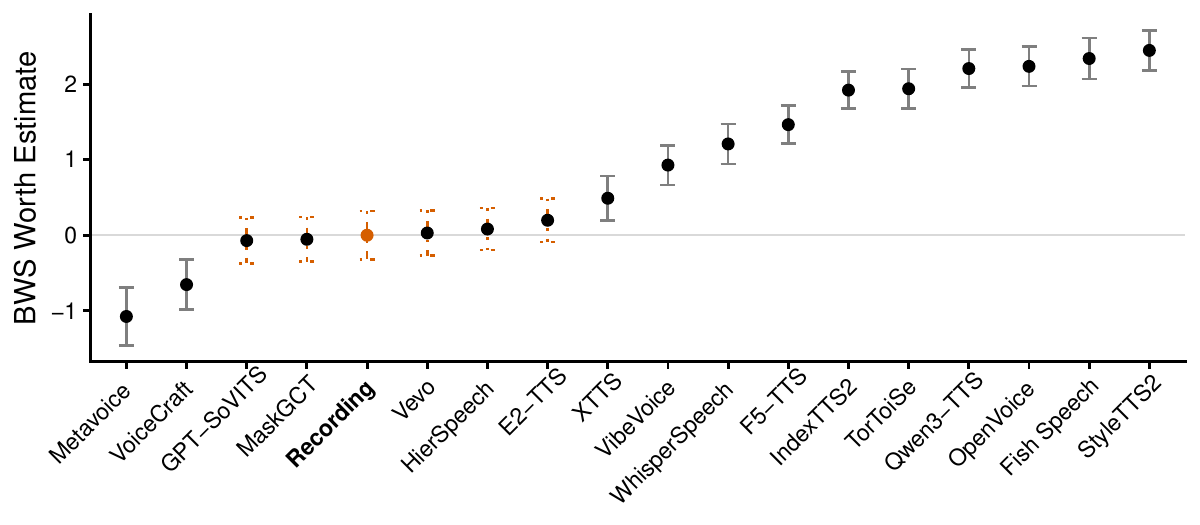}
        \caption{\textsc{intelligibility} (All Speakers)}
    \end{subfigure}
    \hfill
    \begin{subfigure}[b]{0.49\textwidth}
        \centering
        \includegraphics[width=0.925\textwidth]{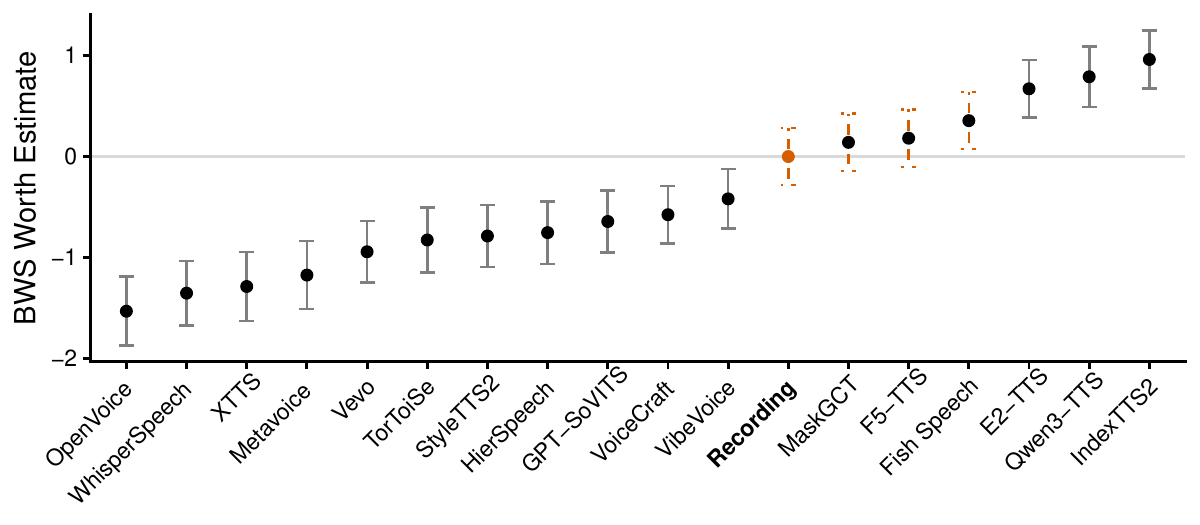}
        \caption{\textsc{reconstruction} (All Speakers)}
    \end{subfigure}
    
    \vspace{0.1em}
    
    \begin{subfigure}[b]{0.49\textwidth}
        \centering
        \includegraphics[width=0.925\textwidth]{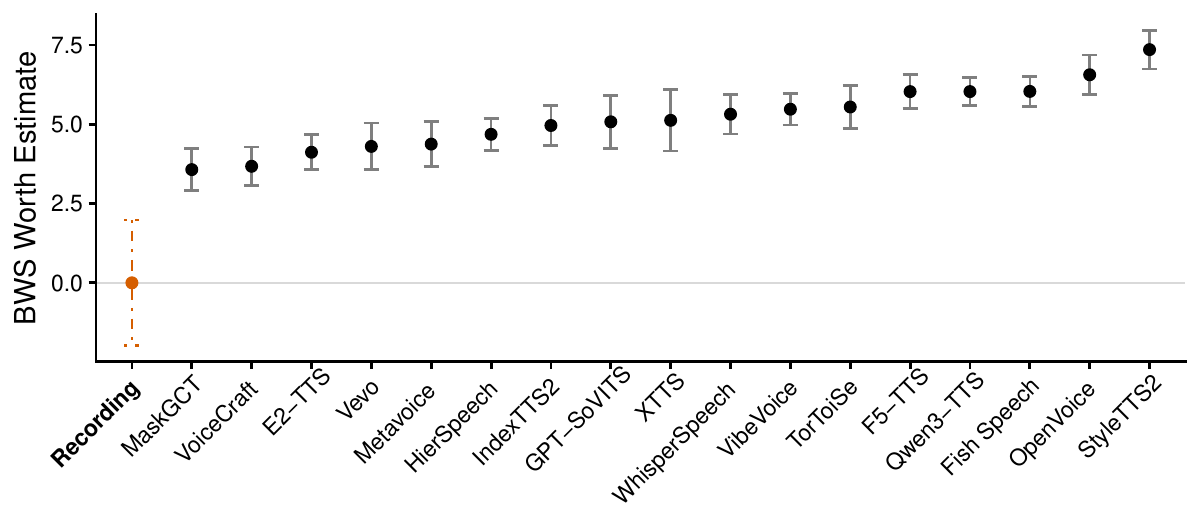}        \caption{\textsc{intelligibility} (Low Intel. Speakers)}
        \label{fig:c}
    \end{subfigure}
    \hfill
    \begin{subfigure}[b]{0.49\textwidth}
        \centering
        \includegraphics[width=0.925\textwidth]{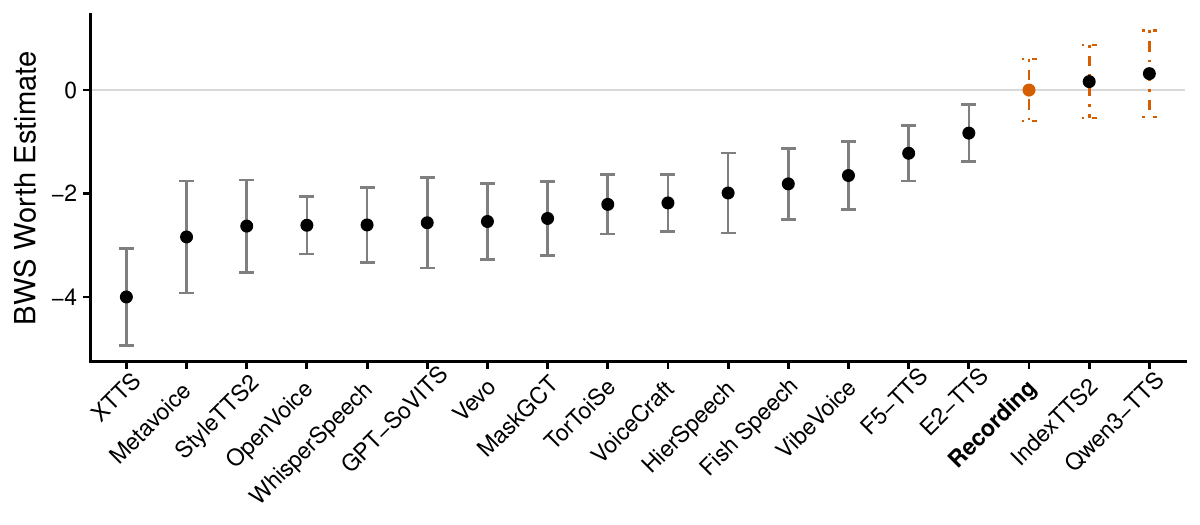}
        \caption{\textsc{reconstruction} (Low Intel. Speakers)}
        \label{fig:d}
    \end{subfigure}
    \vspace{-0.1em}
    \caption{Subjective Best Worst Scaling (BWS) worth estimates with 95\% confidence intervals, with log-worth scores relative to the original recording. Dotted lines represent systems \textbf{without} statistically significant difference to the recording ($p\geq0.05$). Figures (a, b) show all speakers, and (c, d) \textbf{low intelligibility} speakers.}
    \label{fig:subj_bws_combined}
\end{figure*}

\section{Evaluating voice reconstruction}

Voice reconstruction is a specific use-case for TTS which should not be evaluated with general naturalness and similarity evaluations. A number of dimensions are important in voice reconstruction, including for example: (1) the intelligibility of the output, (2) how well the speaker was reconstructed, i.e., evaluating speaker characteristics separated from their speech disorder, (3) accent similarity, and (4) suitability of the output for daily conversation. Here, we focus on only (1) and (2) as the two most important dimensions, leaving (3) and (4) as future work.

To evaluate \textsc{intelligibility} and \textsc{reconstruction}, we need to isolate speaker characteristics from their speech disorder. To do this, we apply situational framing \cite{edlund2024assessing} in our instructions and questions to listeners. We ask them to assess \textsc{intelligibility} without regard to speaker characteristics, i.e., ignoring whether the synthetic speech sounds like the same person. We ask them to assess \textsc{reconstruction} with respect to both intelligibility and speaker characteristics, by comparing the synthetic speech to how they \textit{imagine} the speaker sounded \textit{prior} to developing the condition. Given that listeners are provided with a recording of the speaker \textit{post}-diagnosis, this task is particularly difficult for severely-disordered speech, and of course entirely subjective.

\section{Subjective evaluation}

We chose Best Worst Scaling (BWS) to conduct \textsc{intelligibility} and \textsc{reconstruction} subjective evaluations. BWS has been shown to be a good alternative to MOS which can evaluate system preferences with fewer screens (and therefore fewer listeners) with statistical significance \cite{wells24_interspeech}.

For each type of evaluation, we conducted 5 separate listening tests with different speaker and TTS model combinations. For each screen, 4 systems (including recording) were sampled. Due to the large number of systems, it is not possible to feature all possible 4-system combinations, however all possible pairs are featured at least 5 times. 50 listeners completed each type of evaluation divided across 5 tests each with different sampling. Listeners completed an online Qualtrics survey, were recruited via Prolific, and were paid £4.2 upon completion, according to UK living wage (£12.6 per hour). Listeners currently residing in the United States or the United Kingdom, with no reported hearing impairments, and whose first language is English, were invited to participate. No listener took the test more than once, and each type of evaluation, \textsc{intelligibility} and \textsc{reconstruction}, had different listeners. 

We instructed listeners that they would be assessing synthetic output intended as a personalised communication aid, and that we had used AI to reconstruct how the speaker sounded before they developed the speech impairment. Each screen has 4 audio samples that listeners can select from, plus a reference sample of the target speaker, with \textsc{intelligibility} featuring 25 screens and \textsc{reconstruction} featuring 20 screens in total due to the higher cognitive load of the latter. Attention checks were included in each listening test.
Four listeners were removed from the \textsc{intelligibility} evaluation and three from the \textsc{reconstruction} evaluation, leaving 46 and 47 listeners for analysis, respectively, taking 19 minutes for \textsc{intelligibility} and 22 minutes for \textsc{reconstruction} on average. BWS rankings are computed using a Plackett-Luce model \cite{turner2020modelling}, in line with previous work \cite{wells24_interspeech}.

\begin{table*}[t]
\centering
\caption{Objective and subjective evaluation results for all 17 TTS systems and the original recordings, sorted by Subjective Reconstruction (BWS). $\uparrow$ indicates higher is better, $\downarrow$ indicates lower is better. Best performing TTS systems in each column are in \textbf{bold}.}
\vspace{-0.5em}
\label{tab:main_results}
\resizebox{\textwidth}{!}{%
\begin{tabular}{l rr c ccccccc}
\toprule
\multirow{2}{*}{\textbf{System}} & \multicolumn{2}{c}{\textbf{Subjective (BWS $\uparrow$)}} & \phantom{a} & \multicolumn{7}{c}{\textbf{Objective}}\\
\cmidrule{2-3} \cmidrule{5-11}
& \textsc{Intel.} & \textsc{Recon.} && WER $\downarrow$ & PER $\downarrow$ & UTMOS $\uparrow$ & Spk. Sim. $\uparrow$ & TTSDS\textbar\textit{SAP} $\uparrow$ & TTSDS\textbar\textit{LibriTTS} $\uparrow$ & TTSDS \textit{Mean} $\uparrow$ \\
\midrule
IndexTTS2 \cite{zhou2025indextts2} & 1.921 & \textbf{0.963} &  & 9.9 & 52.4 & 2.88 & 0.670 & 86.9 & 85.3 & 86.1 \\
Qwen3-TTS \cite{hu2026qwen3} & 2.207 & 0.791 &  & 20.0 & 63.9 & 3.58 & 0.573 & 86.4 & 86.5 & \textbf{86.5} \\
E2-TTS \cite{eskimez2024e2} & 0.198 & 0.672 &  & 30.8 & 66.0 & 2.61 & \textbf{0.713} & 90.6 & 81.9 & 86.3 \\
Fish Speech  \cite{liao2024fish} & 2.340 & 0.356 &  & 8.4 & 49.8 & 3.48 & 0.487 & 80.3 & \textbf{86.8} & 83.6 \\
F5-TTS \cite{chen2025f5} & 1.464 & 0.182 &  & 12.3 & 49.4 & 2.94 & 0.619 & 85.8 & 85.2 & 85.5 \\
MaskGCT \cite{wang2024maskgct} & -0.054 & 0.141 &  & 17.5 & 61.7 & 2.58 & 0.627 & \textbf{90.8} & 79.3 & 85.1 \\
\midrule
\textit{Recording (Original)} & \textit{0.000} & \textit{0.000} &  & \textit{20.6} & \textit{65.9} & \textit{2.36} & \textit{0.645} & \textit{93.6} & \textit{77.4} & \textit{85.5} \\
\midrule
VibeVoice \cite{pengvibevoice} & 0.929 & -0.420 &  & 18.6 & 57.5 & 2.75 & 0.486 & 88.7 & 83.0 & 85.8 \\
VoiceCraft \cite{peng2024voicecraft} & -0.654 & -0.577 &  & 35.9 & 72.6 & 2.42 & 0.374 & 88.3 & 78.6 & 83.4 \\
GPT-SoVITS \cite{gptsovits} & -0.071 & -0.644 &  & 27.0 & 70.2 & 2.57 & 0.504 & 87.3 & 82.7 & 85.0 \\
HierSpeech \cite{lee2022hierspeech} & 0.082 & -0.755 &  & 13.7 & 47.6 & 3.62 & 0.496 & 88.9 & 78.6 & 83.7 \\
StyleTTS2 \cite{li2023styletts} & \textbf{2.447} & -0.788 &  & \textbf{6.4} & 48.3 & 3.71 & 0.371 & 80.9 & 85.0 & 83.0 \\
TorToiSe \cite{betker2023better} & 1.940 & -0.828 &  & 15.1 & 48.4 & 3.15 & 0.372 & 82.5 & 85.7 & 84.1 \\
Vevo \cite{zhang2025vevo} & 0.030 & -0.944 &  & 34.8 & 66.2 & 2.67 & 0.563 & 87.1 & 81.2 & 84.2 \\
MetaVoice \cite{metavoice} & -1.076 & -1.176 &  & 43.3 & 76.8 & 2.09 & 0.418 & 80.9 & 77.6 & 79.2 \\
XTTS(v2) \cite{casanova2024xtts} & 0.490 & -1.289 &  & 17.3 & 58.1 & 2.63 & 0.404 & 86.0 & 80.9 & 83.5 \\
WhisperSpeech \cite{whisperspeech} & 1.208 & -1.355 &  & 18.2 & 56.3 & 3.01 & 0.441 & 79.8 & 84.7 & 82.3 \\
OpenVoice \cite{qin2023openvoice} & 2.236 & -1.534 &  & 6.7 & \textbf{45.5} & \textbf{3.72} & 0.283 & 76.8 & 84.1 & 80.4 \\
\bottomrule
\end{tabular}
}
\vspace{-0.5em}
\end{table*}

\begin{table}[t]
\centering
\caption{Spearman rank correlations ($\rho$) between objective measures and subjective BWS estimates for \textsc{intelligibility} and \textsc{reconstruction} by type of speaker.\\ Significance: * $p<0.05$, ** $p<0.01$, *** $p<0.001$.}
\vspace{-0.5em}
\label{tab:correlations}
\resizebox{\columnwidth}{!}{%
\begin{tabular}{l rl rl}
\toprule
\multirow{2}{*}{\textbf{Predictor}} & \multicolumn{2}{c}{\textbf{All Speakers}} & \multicolumn{2}{c}{\textbf{Low Intelligibility}} \\
\cmidrule(lr){2-3} \cmidrule(lr){4-5}
& \textsc{intel.} & \textsc{recon.} & \textsc{intel.} & \textsc{recon.} \\
\midrule
Spk. Sim. & -0.245\phantom{***} & \phantom{-}0.746*** & -0.503*\phantom{**} & \phantom{-}0.610** \\
WER          & -0.802*** & -0.018\phantom{***} & -0.593**\phantom{*} & \phantom{-}0.189\phantom{***} \\
PER          & -0.769*** & \phantom{-}0.117\phantom{***} & -0.622**\phantom{*} & \phantom{-}0.271\phantom{***} \\
UTMOS        & \textbf{0.864}*** & -0.090\phantom{***} & \phantom{-}0.831*** & -0.051\phantom{***} \\
\midrule
TTSDS\textbar\textit{SAP}   & -0.573*\phantom{**} & \phantom{-}0.470*\phantom{**} & -0.752*** & \phantom{-}0.449\phantom{***} \\
TTSDS\textbar\textit{LibriTTS}  & \phantom{-}0.853*** & \phantom{-}0.292\phantom{***} & \phantom{-}\textbf{0.851}*** & \phantom{-}0.185\phantom{***} \\
TTSDS Mean   & \phantom{-}0.034\phantom{***} & \phantom{-}\textbf{0.814}*** & -0.137\phantom{***} & \phantom{-}\textbf{0.734}*** \\
\bottomrule
\end{tabular}
}
\vspace{-1em}
\end{table}

Figure \ref{fig:subj_bws_combined} and Table \ref{tab:main_results} summarise the main results for the \textsc{intelligibility} and \textsc{reconstruction} evaluations with BWS for all speakers in the dataset. For \textsc{intelligibility}, StyleTTS2 ranks the highest, followed by Fish Speech and OpenVoice. As we can observe, most systems rank higher than the recordings, suggesting that, generally, TTS systems are more intelligible than the original speaker. For \textsc{reconstruction}, IndexTTS2 ranks the highest, followed by Qwen3-TTS and E2-TTS. As we can observe, most systems rank lower than the recordings here, suggesting that the synthetic output does not fully capture the desired speaker characteristics in the voice reconstruction task.
Figures \ref{fig:c} and \ref{fig:d} show how TTS systems performs for speakers with low intelligibility ($\text{WER}\geq30$). All TTS systems rank higher than recordings in the \textsc{intelligibility} task. For \textsc{reconstruction}, only IndexTTS2 and Qwen3-TTS rank higher than recordings, suggesting that most TTS systems fail to maintain speaker characteristics while making the speech more intelligible for more severely disordered speech.

\section{Objective evaluation}

We investigate objective measures applied to previous voice reconstruction work by testing whether system rankings according to those measures correlate with the rankings derived from subjective evaluation.
We compute WER by automatically transcribing the synthetic speech and recordings using Whisper \cite{radford2023whisper} (\texttt{turbo} checkpoint) and PER by using the Allosaurus phone recogniser \cite{li2020allosaurus} coupled with transcripts converted to IPA (International Phonetic Alphabet)\footnote{\url{https://pypi.org/project/eng-to-ipa}}. The lack of disordered speech in the Allosaurus training data inflates PER values, however the relative ranking of systems can still serve as a proxy for intelligibility. For MOS prediction, we use the official UTMOS implementation\footnote{\url{https://github.com/sarulab-speech/UTMOS22}}. We use WeSpeaker \cite{wang2023wespeaker} for speaker embedding cosine similarity (Spk.\ Sim.). WER and PER require ground truth transcriptions, while Spk.\ Sim.\ requires a reference signal for comparison, and UTMOS is reference-free \cite{cooper2024review}. There is reason to believe these measures should be able to capture the \textsc{intelligibility} task preferences, however, we hypothesise they are not well-suited for the \textsc{reconstruction} task. WER and PER can serve as a proxy for intelligibility, but listeners might prefer speech which is less intelligible but retains recognisable speaker characteristics to a voice with different identity but perfect intelligibility. Speaker embeddings have been successfully used to recognise dysphonia \cite{aziz2024class}, which shows they do not disentangle speaker identity from speech disorders, decreasing Spk.\ Sim.\ for a successfully reconstructed voice. UTMOS was not exposed to disordered speech in its training data, and has been shown to generalise poorly to new domains \cite{cooper2025progress}.

Distributional evaluation using TTSDS2 was shown to overcome similar shortcomings in other domains \cite{minixhofer2025ttsds2}. We propose that for distributional voice reconstruction evaluation, we can combine the scores of two separate references to predict the trade-off listeners may implicitly weigh. For the high-intelligibility speech reference, we construct a subset of LibriTTS \cite{zen2019libritts} with 145 speakers, one randomly selected utterance from each distinct speaker in the \texttt{test}, \texttt{dev} \texttt{clean} and \texttt{other} splits. We denote the distribution score for this reference as TTSDS\textbar\textit{LibriTTS}. For speaker identity, we use the utterances used to prompt the zero-shot TTS systems, denoting the resulting score as TTSDS\textbar\textit{SAP}. To measure whether a system produces a distribution similar to both reference datasets, we compute the mean of the two, denoted as TTSDS \textit{Mean}.

Table \ref{tab:main_results} presents the objective evaluation results alongside the subjective BWS scores. We observe that different systems excel depending on the specific measure evaluated: StyleTTS2 achieves the lowest WER (6.4\%), indicating high intelligibility, while OpenVoice obtains the highest UTMOS (3.72) and lowest PER (48.3\%). For identity preservation, E2-TTS achieves the highest speaker cosine similarity (0.713), and MaskGCT yields the closest distributional similarity to the disordered reference (TTSDS\textbar\textit{SAP}). However, when balancing intelligibility and identity using the TTSDS \textit{Mean}, Qwen3-TTS, E2-TTS, and IndexTTS2 achieve the highest scores, which are the same three systems that performed best in the subjective \textsc{reconstruction} evaluation.

As shown in Table \ref{tab:correlations}, both WER and PER correlate significantly with \textsc{intelligibility}. UTMOS achieves an even stronger correlation, likely reflecting a bias toward the highly intelligible, typical speech prevalent in its training data. Our proposed distributional measure, TTSDS\textbar\textit{LibriTTS}, achieves a similarly strong correlation with \textsc{intelligibility} across all speakers ($\rho=0.85$, vs. $0.86$ for UTMOS). Notably, it outperforms UTMOS on the low-intelligibility subset ($\rho=0.85$ vs.\ $0.83$).
For \textsc{reconstruction}, speaker similarity shows a strong and significant correlation across all speakers ($\rho=0.75$, $p<0.05$). However, this drops to $\rho=0.61$ for low-intelligibility speakers. Our combined measure, TTSDS \textit{Mean}, outperforms speaker similarity in both cases, achieving $\rho=0.81$ overall and $\rho=0.73$ for the low-intelligibility subset.

\section{Discussion and Conclusion}

With advancements in generative speech synthesis, common subjective evaluation protocols such as MOS naturalness have become saturated, and their existing limitations have been amplified \cite{wells24_interspeech}. Objective evaluation methods sometimes fail to generalise to new domains \cite{cooper2024review} and have to be tested each time a new system or domain is introduced \cite{moller2009quality}.

For \textsc{reconstruction} specifically, listener instructions should be situationally framed: our results show that this leads to a remarkably different outcome than asking them to rate \textsc{intelligibility}. We also show that objective evaluation measures are biased towards intelligibility, and don't capture reconstruction performance. With our application of the distributional TTSDS2 measure to voice reconstruction, we show the trade-off between intelligibility and speaker identity can be quantified objectively for zero-shot TTS -- future work could test if this finding generalises to other datasets and systems.

Some systems, namely Qwen3-TTS \cite{hu2026qwen3}, IndexTTS2 \cite{zhou2025indextts2} and E2-TTS \cite{eskimez2024e2}, consistently outperform the recordings for the full dataset (high and low intelligibility speakers). These systems could be used as baselines for future voice reconstruction work, as they do not need any further adaptation or fine-tuning. However, they underperform for low intelligibility speakers. This could indicate that, as disorder severity increases, zero-shot TTS systems must shift speech further toward a generic high-intelligibility space, and beyond a critical point they can no longer do so while preserving speaker identity.

Our subjective evaluation approach can be extended to other evaluation paradigms, such as in-context usability, adequacy, prosody or accent. This framework can still be used to assess personalised TTS preferences by a VOCA user instead of relying on listeners which are not ultimately going to use the device. Ultimately, future work should aim to expand on user-centric approaches when evaluating voice reconstruction.

\section{Acknowledgements}
This study has been approved by the School of Informatics Ethics' Committee at the University of Edinburgh, with reference number 997684. The first author was supported by the UKRI Centre for Doctoral Training in Natural Language Processing, funded by UKRI (grant EP/S022481/1). We want to thank Mark Hasegawa-Jonhson and collaborators for creating and sharing the Speech Accessibility Project (SAP) dataset, and to the SAP participants for donating their voices.

\section{Generative AI use disclosure}
We did not use any generative AI for this work, except to generate the synthetic speech stimuli and parts of the demo page.

\bibliographystyle{IEEEtran}
\bibliography{mybib}

\end{document}